\documentstyle[aps,prl,graphicx]{revtex} 
\begin{document}

\title{Electronic friction and liquid-flow-induced voltage in nanotubes }
\author{B.N.J. Persson$^{1,2,3}$, U. Tartaglino$^{4}$,  E. Tosatti$^{3,4,5}$ and H. Ueba$^6$}

\vskip 0.3cm

\address{$^1$ IFF, FZ-J\"ulich, 52425 J\"ulich, Germany}

\address{$^2$Kavli Institute for Theoretical Physics, University of California,
Santa Barbara, CA 93106-4030, USA}

\address{$^3$ International Centre for Theoretical Physics (ICTP),
P.O. Box 586, I-34014 Trieste, Italy}

\address{$^4$ International School for Advanced Studies (SISSA), and
INFM Democritos National Simulation Center, 
Via Beirut 2, I-34014 Trieste, Italy}
 
\address{$^5$ Laboratoire de Mineralogie-Cristallographie de Paris,
Universite' Pierre et Marie Curie, 4 place Jussieu, 75252 Paris 
Cedex, France}

\address{$^6$ Department of Electronics, Toyama University, Gofuku, Toyama,
930-8555, Japan}

\maketitle

\begin{abstract}
A recent exciting experiment by Ghosh {\em et al}\cite{flow} reported 
that the flow of an ion-containing  liquid such as water through  
bundles of single-walled carbon nanotubes induces a voltage in 
the nanotubes that grows logarithmically with the 
flow velocity $v_0$. We propose an explanation for this observation. 
Assuming that the liquid molecules 
nearest the nanotube form a 2D solid-like monolayer pinned through
the adsorbed ions to the nanotubes, the monolayer sliding will
occur by elastic loading followed by local yield (stick-slip). 
The drifting adsorbed ions produce a voltage
in the nanotube through electronic friction against free electrons 
inside the nanotube. Thermally excited jumps over force-biased 
barriers, well-known in stick-slip, can explain the logarithmic voltage 
growth with flow velocity. We estimate the short circuit current
and the internal resistance of the nanotube voltage generator.
\end{abstract}
\vskip 0.5cm


%%%%%%%%%%%%%% main text %%%%%%%%%%%%%%%%

\vskip 1cm

{\bf 1. Introduction}

In a recent, remarkable experiment\cite{flow}, it was observed that the flow
of an ion-rich liquid such as water through  bundles of single-walled carbon nanotubes 
induces a {\em voltage} in the nanotube sample along the direction 
of the flow. Strikingly, the voltage grows logarithmically with the liquid 
flow velocity over nearly six decades. The voltage magnitude and 
sign depend sensitively on the ionic conductivity and 
on the polar nature of the liquid, suggesting that ions adsorbed, 
or located in the immediate vicinity of the carbon nanotubes must be
involved in its generation. This work highlights 
the device potential for nanotubes as sensitive flow sensors; it also
presents very interesting and intriguing questions about the underlying physics. 

As an explanation for the observed effect, Ghosh et al\cite{flow} suggested that 
the dominant mechanism responsible for this highly nonlinear response to
the flow could involve a direct forcing of the free electrons in the nanotubes 
by the fluctuating Coulombic field of the liquid flowing past the nanotubes,
through pulsating asymmetric ratchets. The model assumes adsorbed ions
which nonetheless do not undergo any average physical drift, and needs
to invoke ratchets for which there seems to be no independent evidence.
Earlier work\cite{Kral} had suggested instead that molecular layers 
of liquid coating the nanotube physically slip along its surface, and excite 
a phonon wind which drags the free carriers in the tube.
However, this model would yield an induced voltage linear in the
fluid flow velocity. Back in the 19th century, Helmholtz\cite{Helm} had
proposed an electrokinetic mechanism where a voltage induced by a liquid
flowing past a solid substrate will appear as a streaming potential involving the 
ions carried by fluid flow in the diffuse (Debye)layer at the interface, 
while the mobile charge carriers in the substrate play no role. 
This mechanism too would, for small flow velocities, predict 
a linear voltage increase with flow velocity. 
Our purpose here will be to propose an alternative model capable of explaining 
the observations -- and of making prediction that could be checked by 
further work -- based on flow-induced ion drift. 

\begin{figure}[htb]
\begin{center}
\includegraphics[width=0.90\textwidth]{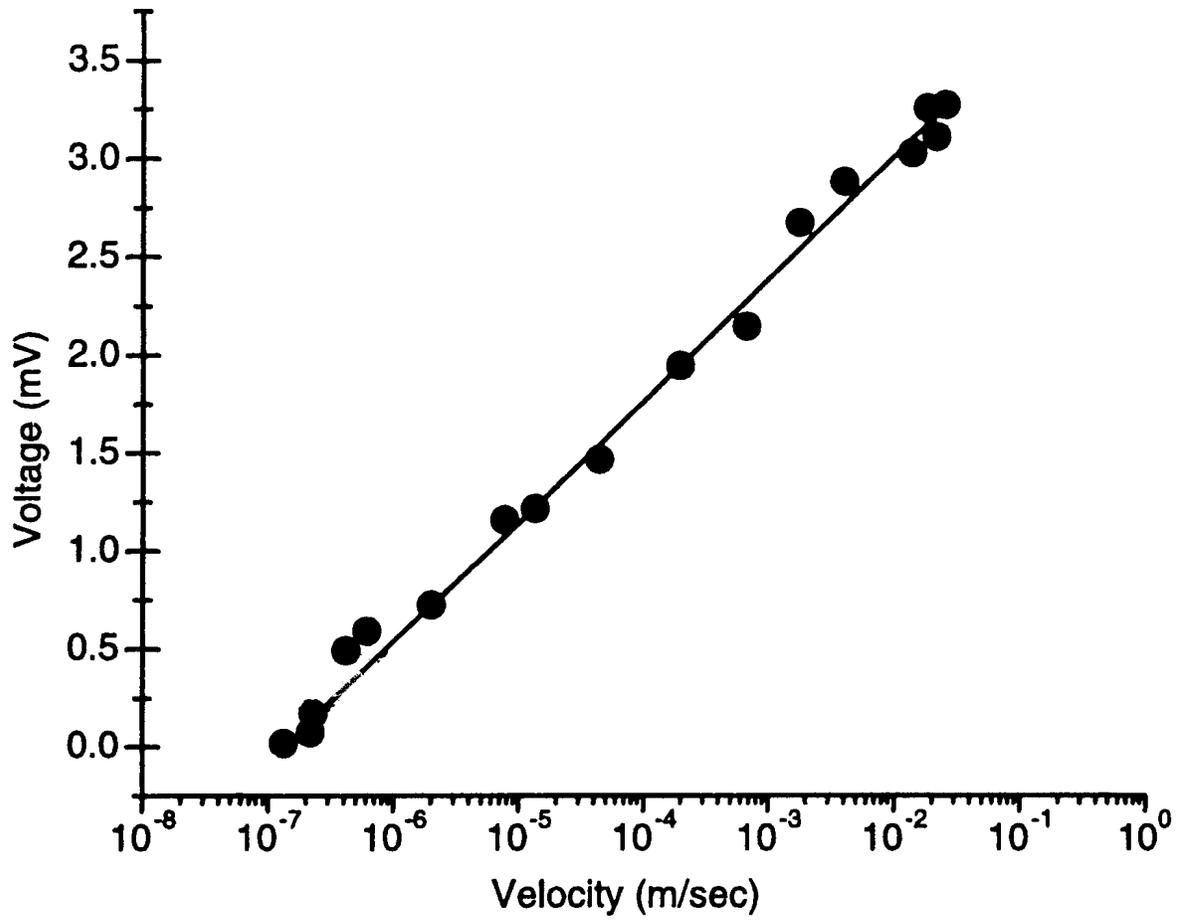}
\end{center}
\caption{
The voltage $U$ as a function of the logarithm of the (average) fluid 
velocity $v_0$. The solid line is a fit of the form $U=U_c \ {\rm ln} (v_0/v_c)$.
Adapted from [1].
} 
\label{logcreep}
\end{figure}

The experiments where performed on nanotube bundles (the nanotubes
are not aligned).
The nanotubes are are micrometer sized in length and nanometric in diameter.
Fig. \ref{logcreep} shows the induced voltage as a function of the 
logarithm of the water (average) flow velocity $v_0$. 
The solid line is a fit to the 
experimental data given by
$$U=U_c \ {\rm ln}\left ({v_0\over v_c} \right )$$
where $U_c = 0.26 \ {\rm mV}$ and $v_c=1.54\times 10^{-7} \ {\rm m/s}$.  

\begin{figure}[htb]
\begin{center}
\includegraphics[width=0.90\textwidth]{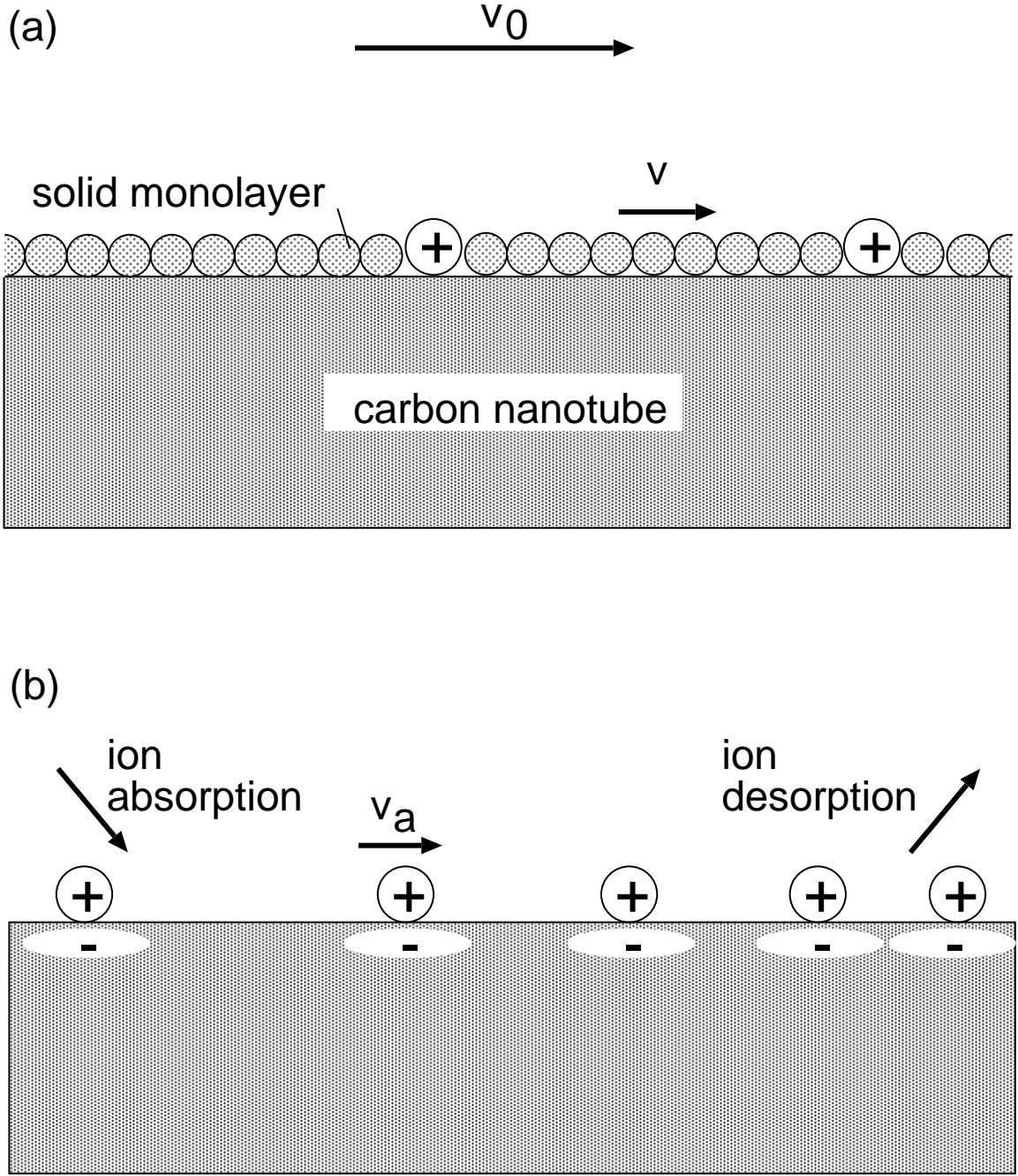}
\end{center}
\caption{Carbon nanotube immersed in a flowing liquid. We assume that 
the liquid molecules nearest the nanotube form a 2D solid-like monolayer, 
pinned to the nanotube by the adsorbed ions. As the liquid flows, the 
solid sliding motion along the nanotube  
is of a stick-slip nature: time intervals of elastic deformation 
(loading phase) are followed by rapid local yield events. 
This will result in a force on the ions which increases logarithmically
with the fluid flow velocity. The drifting 
adsorbed ions will produce a voltage in the nanotube through electronic 
friction against free electrons inside the nanotube. 
In the stationary state there is a net flow of ions 
adsorbing on the nanotube upstream, and a net flow of ion desorption 
downstream.}
\label{Pol}
\end{figure}
 
\vskip 0.5cm

{\bf 2. Ion stick-slip model}

Although the reasoning is general, let us focus here on water as 
the fluid, and start by assuming that ions dissolved in it become 
adsorbed on the carbon nanotube surfaces, see Fig. \ref{Pol}. We 
also postulate that the water molecules in the first layer around 
a nanotube form an (incommensurate) solid-like monolayer. Recent 
simulations provided evidence for instances where that does occur
\cite{layer}. This 2D solid-like water layer will 
experience a pinning potential on the nanotube, that is mediated 
through the adsorbed ions, and is in turn dragged or pulled forward 
by the external liquid water flow. The solid sliding motion along 
the nanotube occurs by stick-slip.
Time intervals of elastic deformation (loading phase) are 
followed by rapid local yield events. During the loading phase the 
local stress in the 2D-solid near an ion will increase linear with time, until 
a critical stress $\sigma_{\rm c}$ is reached, when a local 
yield (rapid rearrangement process) takes place. We denote 
the elementary solid element where most of the rearrangement occur 
as the ``stress block".

A linear (or nearly linear) relationship between driving force and the 
logarithm of the induced drift velocity has been observed for many
systems exhibiting local stick-slip motion, e.g., charge density 
wave and flux line systems\cite{earlywork}, as well as for atomic 
force microscope tips moving in a substrate pinning potential\cite{tip}. 
These results are collectively understood through a standard 
picture involving thermally excited jumps over force-biased 
barriers\cite{book}. This picture will apply for all systems which 
can be described as an effective elastic solids pinned by defects, 
and driven by an external force. The basic picture consist of elastic 
loading, followed by rapid, local jumps over pinning barriers; in 
these jumps local regions (stress domains) of the elastic solid move 
forward. When thermal activation of such a stick-slip 
process is taken into account, 
the average stress at yield increases logarithmically with the 
velocity of the dragging fluid.

Assume that the (average) velocity $v$ of the adsorbed 2D solid-like 
water monolayer depends linearly on the (average) fluid flow 
velocity $v_0$, $v=\alpha v_0$, 
as expected for a Newtonian liquid. During the loading phase the strain in 
a stress block will be of order $u/L$, 
where $L$ is the linear size of a stress block and 
$u$ the displacement $u=vt$. The local shear stress in the stress block will 
be of order $\sigma\approx E u/L$, where $E$ is the elastic modulus
of the 2D solid. We define the loading force $F=La \sigma 
= Eau = ku$, where $a$ is the thickness of the 2D solid monolayer film, and
the effective spring constant is $k=Ea$. We could expect $E$ to be similar 
to the elastic modulus of ice, i.e., of order $5\times 10^9 \ {\rm Pa}$
and with $a\approx 0.3 \ {\rm nm}$ we get $k \approx 1.5 \ {\rm N/m}$.
Assuming the critical strain at yield to be of order $0.1$ we get
the critical displacement $u_{\rm c} \approx 1 \ {\rm \AA}$ and the elastic barrier
$\epsilon = k u_{\rm c}^2 /2 \approx 0.05 \ {\rm eV}$, i.e.,
in the same range of energy as a hydrogen bond.  

\begin{figure}[htb]
\begin{center}
\includegraphics[width=0.70\textwidth]{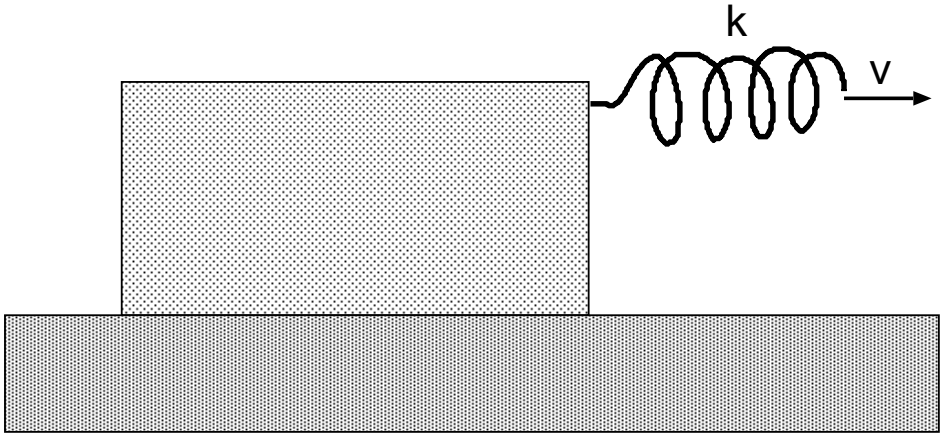}
\end{center}
\caption{
Mechanical model representing the stick-slip (loading by elastic deformation 
followed by abrupt yield) of the 2D-solid in the vicinity of an ion 
[see Fig. \ref{Pol}(a)]. The spring 
represent the elastic deformation of the 2D solid. When the
spring force reaches a critical value (the static friction force)
the block start to slip, reverting back to the pinned state
when the spring force approximately vanishes.} 
\label{Model}
\end{figure}

The picture presented above is represented in Fig. \ref{Model},
where the spring represents the elastic deformation of the 2D solid. When the
spring force reaches a critical value (the static friction force)
the block starts to slip and will revert back to the pinned state
when the spring force approximately vanishes. This fast slip event
corresponds to a yield or rearrangement process of the
2D solid in the vicinity of the ion. 
In \cite{electronicfriction,stickslip} we studied the relationship between the drive velocity $v$
and the (average) spring force $F$ for the model in Fig. \ref{Model} 
including temperature effects. 
If $F_{\rm c}$ denotes the ``static'' friction force (the spring force necessary
in order to start the slip of the block), at temperatures $T > 0 \ {\rm K}$
thermal excitation will allow the block to depin before the spring force
has reached the critical value $F_{\rm c}$.
The rate of thermal excitation over a barrier $\epsilon$ can be assumed
to be of the standard form
$$ w = \nu e^{-\beta\epsilon'} $$
where the prefactor $\nu\approx 10^{10}\,\mathrm{s}^{-1}$ and the effective
depinning barrier $\epsilon'$ ($0<\epsilon'<\epsilon$) depends on the
local shear stress in the contact area. By considering the statistical
distribution of the microscopic shear stress, the friction versus speed
can be calculated\cite{electronicfriction}.
Three regimes are found, depending on the magnitude of the speed.
We define the parameter $v^*$:
$$ v^* = \frac{F_{c}\nu}{2 k\beta\epsilon} e^{-2\beta\epsilon} \eqno(1) $$
In a wide intermediate range of speeds, where
$1 << v/v^* << e^{2\beta\epsilon}$, we get
$$ F = \frac{F_{c}}{4\beta\epsilon} \ln\left(\frac{v}{v^*}\right)
   \eqno(2) $$
In the extreme low velocity limit, where $v/v^* << 1$,
we have $$F =  {k v \over \nu} e^{\beta \epsilon}$$
For very high velocity instead, $v/v^* >> {\rm exp} (2 \beta \epsilon)$, 
we have on the other hand
$$F = {F_{\rm c} \over 2}\left (1-{v^*\over 2 \beta \epsilon v} e^{2\beta \epsilon} \right )$$

We now assume that the force $F$ will also act on the ion which 
is at the same time embedded in the 2D solid monolayer and adsorbed on the
nanotube. This force will result in a slow average drift motion of the 
ion along the water flaw direction. Since $F$ is very small, linear response 
should apply. The adsorbed ion drift velocity is thus $$v_a=\mu F\eqno(3)$$ 
where the mobility $\mu$ depends on the lateral ion diffusion barriers.
Combining (2) and (3) gives
$$v_a = v_a^c \ 
{\rm ln} \left ( {v_0\over v_c}\right ),\eqno(4a)$$
where $v_c = v^* / \alpha$ and 
$$v_a^c = {\mu F_{\rm c}\over 4 \beta \epsilon}.\eqno(4b)$$

If we write $F_{\rm c} = k u_{\rm c}$, where $u_{\rm c}$ is
the displacement necessary in order induce the yield,
we get from (1)
$$v_c = v^* / \alpha =
{u_{\rm c} \nu \over 2 \alpha \beta \epsilon} e^{-2\beta \epsilon}\eqno(5)$$
Note that (5) can be written as
$$2 \beta \epsilon = {\rm ln} \left ({u_c \nu \over 2 \alpha\beta \epsilon v_c } \right )\eqno(6)$$
which shows that $\epsilon$ is very insensitive to the exact value of $\alpha$, $u_c$ and
$\nu$. If we assume 
$\alpha = 0.1$, $u_c \approx 1 \ {\rm \AA}$ and $\nu \approx 10^{10} \ {\rm s}^{-1}$
we get with (from Fig. \ref{logcreep}) $v_{\rm c} = 1.54 \times 10^{-7} \ {\rm m/s}$, 
$\epsilon \approx 0.2 \ {\rm eV}$ which is very plausible.

\vskip 0.5cm

{\bf 3. Flow induced current and voltage: electrical properties}

Assume that there is a 2D concentration $n_e$ of conduction electrons in the
carbon nanotube. We shall use here a semi-classical picture for the motion 
of the electrons.  Under the ion-originated external forces the conduction electrons 
drift with the velocity $v_e$, so that the electric current in a nanotube 
will be $I = 2 \pi r n_e e v_e$, where $r$ is the radius of the
nanotube. The equation of motion for the drift velocity is given by
$$m_e \dot v_e = - m_e {1\over \tau} v_e - m_a \eta (v_e-v_a) {n_a \over n_e} + e E\eqno(7)$$
where $\tau$ is a Drude relaxation time, $\eta$ an {\it electronic} friction
coefficient, $n_a$ the concentration of adsorbed ions, $E$ the electric field 
in the 2D nanotube, and $v_a$ is the drift velocity of the adsorbed ions.
In deriving (7) we assumed that the frictional drag from the moving 
adsorbates to be distributed uniformly over all the conduction electrons in 
the nanotube. The electronic friction coefficient
$\eta$ can be related to the adsorbate-induced increase in the nanotube resistivity.\cite{book}
Thus, when $v_a =0$ (no fluid flow) we get in steady state
$$v_e= {e\tau E/m_e \over 1+\eta \tau (m_a/m_e)(n_a/n_e)}$$
so that the current
$$I=  
{2\pi r  n_e e^2 \tau E/m_e \over 1+\eta \tau (m_a/m_e)(n_a/n_e)}$$
Thus, $U=RI$, where the resistivity
$$R= 
R_0\left (1+\eta \tau {m_a \over m_e}{ n_a \over n_e}\right )=R_0(1+\kappa)$$
where $R_0$ is the nanotube resistivity in the absence of adsorbates:
$$R_0 = 
L m_e /( 2\pi r  n_e e^2 \tau)$$

We consider now the full equation (7) in two different limiting cases,
namely short circuit and open circuit. In the first case $E=0$ so that
in the stationary case (1) takes the form 
$$ 0 = - m_e {1\over \tau} v_e - m_a \eta (v_e-v_a) {n_a \over n_e}$$
so that
$$ v_e= {(m_a/m_e) (n_a/n_e) \eta \tau v_a \over 1+(n_a/n_e)(m_a/m_e) \eta \tau}$$
and thus  
$$ I= {2 \pi r n_a (m_a/m_e) \eta \tau e v_a \over 1+\kappa}
= {\kappa \over 1+\kappa}2\pi r n_e e v_a \eqno(8)$$ 
The second limiting case (open circuit) correspond to $v_e=0$ so that, from (7),
$U=LE$ is given by
$$U= 
{L m_a \eta \over e} {n_a \over n_e} v_a = \kappa R_0 2\pi r n_e e v_a\eqno(9)$$
We expect $\kappa > 1$ so that for the short circuit case the current
$I \approx 2\pi r n_e e v_a$ is nearly equal to the current expected if the
nanotube conduction electrons drift with the same velocity as the adsorbed
ions.

We note that $R$ given above represents the {\em internal resistance}
of the nanotube bundle as an EMF generator.
The measured resistivity $\rho_0$ of the nanotube bundles in the normal atmosphere
is $\rho_0 = 0.02 \ {\rm \Omega m}$ so that the resistance $R_0 = \rho_0 L/A$ where
the length $L= 1 \ {\rm mm}$ and the cross section area $A=0.4 \ {\rm mm}^2$. Thus
$R_0 = 50 \ {\rm \Omega}$. If the electron flow velocity $v_e$ is measured in 
${\rm m/s}$ and if the electron concentration $n_e = 2\times 10^{27} \ {\rm m}^{-3}$ we get
the electron current in Ampere,
$$I= An_e e v_e \approx 100 \ v_e,$$
and the induced voltage in ${\rm mV}$,
$$U = R_0 I = 5\times 10^6 \ v_e.\eqno(10)$$
The lowest flow velocity for which a flow induced voltage was measured 
was $2\times 10^{-7} \ {\rm m/s}$  and in that case the induced voltage was 
about $0.2 \ {\rm mV}$. Now, this is 
basically the value one would obtain if the drift velocity
of the electrons in the nanotubes equaled the (average) liquid flow
velocity and if the resistivity $R$ for the liquid-immersed nanotube bundles 
were the same as that of the bundles in the atmosphere. Thus 
with $v_e=2\times 10^{-7} \ {\rm m/s}$ Eq. (10) gives $$U= 1 \ {\rm mV}.$$
This is only five times higher than the measured voltage indicating 
(if as we expect $R$ is similar to $R_0$) that the electron drift 
velocity in the nanotubes, in the limiting case
of the lowest possible fluid flow velocity where the induced voltage can be observed,
is of order the average fluid flow velocity. At higher fluid flow velocity 
drag is less perfect, and this is no longer the case, since the voltage 
increases logarithmically with the fluid flow velocity. At the highest
studied fluid flow velocity $v_0 \approx 10^{-2} \ {\rm m/s}$ 
the electron drift velocity would be roughly a factor of $10^5$ lower 
than the average fluid flow velocity.

\vskip 0.5cm

{\bf 4. Consequences of the model}

Because in our stick-slip model the logarithmic voltage increase with flow
velocity is due to thermally activated jump of the ions, biased 
by the force arising from the drift motion of the solid monolayer, we can
anticipate first of all a strong temperature dependence. This is
apparent in Eq.  (4) above, where the strongest temperature dependence is derived
from the thermal Boltzmann factor which occurs in the microscopic expression 
for the mobility $\mu$, that should thus increase exponentially with
temperature. In addition there is also a linear temperature 
prefactor $1/\beta \sim T$, but this is in most cases negligible 
compared to the Boltzmann factor. 
There may also be a temperature dependence arising from a temperature
dependence of the concentration of adsorbed
ions. In any case, a strong temperature 
dependence of the induced voltage 
is predicted by our model and should provide a first possibility
of experimental test. 

A second aspect, which strongly distinguishes our model from fluctuating
ratchets, is the flow-induced concentration of ions from upstream to
downstream. In the free fluid, every ion is neutralized by some counter-ion
of opposite charge. When an ion, say positive, adheres upstream to the nanotube 
surface, it will get to a large extent neutralized by the image electron 
charge in the nanotube. The corresponding negative counter-ion in the fluid 
is no longer neutralized. Being made redundant by the positive ion adsorption, 
it must go somewhere else. When on the other hand, after having drifted along
a nanotube the positive ion leaves the nanotube surface some distance 
downstream, it will lose its neutralizing image electron, and must at 
that point again recuperate a counter-ion.
This suggests that the flow-induced drag of positive adsorbed ions
will be accompanied by the simultaneous flow of image electron inside the
nanotube, {\em and} by a drift of negative counter-ions inside the fluid,
that will migrate from upstream, where they are redundant, to downstream, 
where they are needed. In that case we have two negative currents, one 
electronic inside the nanotube and one ionic, due to the negative counterions, 
against only one ionic positive current, that of the adsorbed ions. The 
end result is a net negative current, as observed. 

This outcome  differs from that of models such as that by Kumar et al, 
where no physical drift of ions is involved. If, as we suggest, ionic 
transport is involved, then there will not generally be a spatially 
uniform steady state. To be sure, there could be a non-uniform steady 
state where the ion concentration (of both signs) is slightly higher 
downstream than upstream. Depending on geometrical conditions, flow 
velocity etc, the ions might or might not have the time to diffuse 
backwards upstream, against the flow, so as to establish such a steady 
state equilibrium. That suggests the possibility to look in principle 
at that {\em ion concentration gradient} as the possible signature of 
an ion drift mechanism, although it is at the moment unclear whether 
such a gradient could in fact be measured. 

A third consequence of our model is the presence of characteristic 
stick slip {\em noise}, that should be readily observable by frequency 
analysis of the induced voltage. Straightforward as this seems, it
might constitute the most direct and simplest test of the model proposed.

\vskip 0.5cm

{\bf 5. Summary and conclusions}

We propose that the logarithmic voltage observed in nanotubes
bundles upon immersion in a flowing liquid with ions dissolved,
could have the same thermal barrier jump origin as that observed 
for frictional stick-slip processes between solids. Once ions
adsorb on the nanotubes, they should be dragged not just by flowing 
individual liquid molecules, but by some kind of solid layer 
of molecules which surrounds the nanotube, and which is rigid 
enough to support the stick-slip. Each drifting adsorbed ion can 
in turn drag along electrons inside the nanotube, through the very 
same electronic frictional force which enables a fixed adsorbed 
ion to cause electrical resistance inside the nanotube. 

By assuming this mechanism to be at work we obtained very simple 
formula for the nanotube generator electrical properties. Expressions
for the internal resistance, short circuit current and open circuit 
voltage obtained in this manner seem entirely reasonable. Moreover
the latter quantities increase logarithmically with the flow velocity,
as is seen experimentally. 

Various tests can be considered for our mechanism. 
First, the logarithmic voltage increase should be strongly 
temperature dependent. Secondly, the physical
dragging of adsorbed ions should cause the ion concentrations in the fluid
to become non-uniform, and in particular to become higher downstream.
Third, a characteristic stick-slip noise spectrum should arise
on top of the DC flow-induced voltage. It is hoped that these preliminary 
simple predictions will stimulate further experimental work on this intriguing effect.

\vskip 0.5cm

{\bf Acknowledgments} We are much indebted to Professors N. Kumar and A.K. Sood
for introducing us to their new effect, for making unpublished information
available to us, and for illuminating discussions. Work at SISSA was  sponsored 
through INFM PRA NANORUB, through MIUR COFIN 2003 as well as  FIRB RBAU01LX5H and 
FIRB RBAU017S8R, and by Regione Friuli Venezia Giulia. One of us (BNJP) is grateful
to the International Centre for Theoretical Physics, where this work was 
initiated, for hospitality. 
BNJP also thank the EC for a ``Smart Quasicrystals''
grant under the EC Program ``Promoting Competitive and Sustainable Growth''.
This research was supported by the National Science Foundation under
Grant No. PHY99-07949.

\end{document}